# Indicators of retention in remote digital health studies: A cross-study evaluation of 100,000 participants

Abhishek Pratap, MS[1,2]; Elias Chaibub Neto, PhD[1]; Phil Snyder, BS[1]; Carl Stepnowsky, PhD[3,4]; Noémie Elhadad, PhD[5]; Daniel Grant, PhD[6]; Matthew H. Mohebbi, BS[7], Sean Mooney, PhD[2]; Christine Suver, PhD[1]; John Wilbanks, BA[1]; Lara Mangravite, PhD[1]; Patrick Heagerty, PhD[8]; Pat Arean, PhD[9]; Larsson Omberg, PhD[1]

[1] Sage Bionetworks, Seattle, WA, USA
[2] Department of Biomedical Informatics and Medical Education, University of Washington, Seattle, WA, USA
[3] University of California, San Diego, CA, USA
[4] American Sleep Apnea Association, Washington, DC, USA
[5] Columbia University, New York, USA
[6] Novartis Pharmaceutical Corporation, East Hanover, NJ, USA
[7] GoodRx, Santa Monica, CA, USA
[8] Department of Biostatistics, University of Washington, Seattle, WA, USA
[9] Department of Psychiatry & Behavioral Sciences, University of Washington, Seattle, WA, USA

# Abstract

Digital technologies such as smartphones are transforming the way scientists conduct biomedical research using real-world data. Several remotely-conducted studies have recruited thousands of participants over a span of a few months. Unfortunately, these studies are hampered by substantial participant attrition, calling into question the representativeness of the collected data including generalizability of findings from these studies. We report the challenges in retention and recruitment in eight remote digital health studies comprising over 100,000 participants who participated for more than 850,000 days, completing close to 3.5 million remote health evaluations. Survival modeling surfaced several factors significantly associated(P < 1e-16) with increase in median retention time **i)** Clinician referral(increase of 40 days), **ii)** Effect of compensation (22 days), **iii)** Clinical conditions of interest to the study (7 days) and **iv)** Older adults(4 days). Additionally, four distinct patterns of daily app usage behavior that were also associated(P < 1e-10) with participant demographics were identified. Most studies were not able to recruit a representative sample, either demographically or regionally. Combined together these findings can help inform recruitment and retention strategies to enable equitable participation of populations in future digital health research.

## Introduction

Traditional in-person clinical trials serve as the cornerstone of modern healthcare advancement. While a pivotal source of evidence generation for advancing clinical knowledge, in-person trials are also costly and time-consuming, typically running for 3-5 years from conception to completion, at a cost of millions of dollars per study. These timelines have often meant that promising treatments take years to get to market, which can create unnecessary delays in advancing clinical practice. Additionally, clinical research suffers from several other challenges[1,2] including 1) recruiting sufficiently large and diverse cohorts quickly, and 2) tracking day-to-day fluctuations in disease severity that often go undetected in episodic in-clinic evaluations[3,4]. Scientists have recently turned to digital technology[5,6] to address these challenges, hoping to collect real-world evidence[7] from large and diverse populations to track long-term health outcomes and variations in disease trajectories at a fraction of the cost of traditional research[8].

The global penetration[9] and high-frequency usage of smartphones (up to 4 hours daily[10,11]) offer researchers a cost-effective means to recruit a large number of participants into health research across the US (and the world)[12,13]. In the last 5 years, researchers have conducted several large scale studies[14–22] including deploying interventions[23,24] and running clinical trials[25–27] using mobile technologies. These studies are able to recruit at-scale because participants can be identified and consented[28] to participate in the study without ever having stepped foot in a research lab, with significantly lower costs than conventional clinical trials[23,24]. Mobile technologies also allow investigators an opportunity to collect data in real-time based on people's daily lived experiences of the disease, that is, real-world data[7]. Rather than retrospectively asking people to recall their health over the past week or month, researchers using mobile technologies can assess participants frequently including outside clinic and at important points in time without having to rely on recall that is known to have bias[29]. While these studies show the utility of mobile technology, challenges in participant diversity and long-term participant retention still remain a problem[30].

Digital studies continue to suffer from long-term participant retention problems that also plagued internet-based studies[31,32] in the early 2000s[33–35]. However, our understanding of factors impacting retention in remote research remains to be limited. High levels of user attrition combined with variations in long-term app usage may result in the creation of a cohort that may not represent the population of interest in regard to demographics, disease status, and disability. This has called into question the reliability and utility of the collected data from these studies[36]. Furthermore, while for many digital health studies, anyone eligible can self-select to join, this broad "open enrollment" recruitment model may be prone to selection and ascertainment bias[36]. Systematic evaluation of participant recruitment and retention could help detect such confounding characteristics that may be present in large scale remotely collected data and has been shown to severely impact the generalizability of the derived statistical inference[36,37]. Participant retention may also be partially dependent on the engagement strategies used in remote research. While most studies assume participants will remain in a study for altruistic reasons[38], other studies provide compensation for participant time[39], leverage partnerships with local community organizations, clinical registries, and clinicians to encourage participation[23,24]. Although monetary incentives are known to increase participation in research[40], we know little about the relative impact of demographics, recruitment and different engagement strategies on participant retention, especially in remote health research.

The purpose of this study is to document the drivers of retention, and long term study app usage in remote research. To investigate these questions we have compiled user engagement data from eight digital health studies that enrolled more than 100,000 participants from throughout the US between 2014-2019. These studies assessed different disease areas including asthma, endometriosis, heart disease, depression, sleep health, neurological diseases and consisted of a combination of longitudinal subjective surveys and objective sensor-based tasks including passive data[41] collection. The diversity of the collected data allows for a broad investigation of different participant characteristics and engagement strategies that may be associated with higher retention including assessment of representational bias in the collected real-world data.

## Results

**Participant Characteristics**

The combined user activity data from eight digital health studies resulted in a pool of 109,914 participants who together completed approximately 3.5 million tasks on more than 850,000 days (Table 1). Across the studies, the majority (Median=65.2%) of participants were between 17-40 years with those 60 years and older being the least represented (Median across studies=6% of the study population). The sample had a larger proportion of Females (Median=56.9%) however it varied significantly across the studies (Range=29.4-100%). A majority of recruited participants were Non-Hispanic Whites (Median=75.3%) followed by Hispanic/Latinos (Median=8.21%) and African-American/Blacks (Median=3.45%) (Table 2). With the exception of the Brighten study, the race/ethnic diversity of the sample also showed a marked difference from the 2010 census data. Minority groups were under-represented in the present sample with Hispanic/Latinos and African-America/Black showing a substantial difference of -8.09% and -9.15% respectively compared to the 2010 census metrics (Table 2, Figure 1-b). Across the studies, the median proportion of recruited participants per state showed notable differences from the state's population proportion of the US (Figure 1.a).

**Participant Retention**

As is the nature of these studies, participants were required to complete all health assessments and other study-related tasks (eg: treatments) through a mobile application (app) throughout the length of the study. The median time participants engaged in the study in the first 12 weeks was 5.5 days of which in-app tasks were performed on 2 days (Table 2). Higher proportions of active tasks were completed by participants during the evening(4-8 PM) and night(8-12 Midnight) hours (Figure 2-a). Across the studies, the median retention time varied significantly ($P < 1e-16$) between 2 and 12 days with the Brighten study being an outlier with a higher median retention of 26 days (Figure 2-b). A notable increase in median retention time was seen for sub-cohorts that continue to engage with the study apps after day one and beyond (Figure 2-c). For example, the median retention increased by 25 days for the sub-cohort that was engaged for the first 8 days. The participant retention also showed a significant association with participant characteristics. While older participants (60 years and above) were the smallest proportion of the sample, they remained in the study for a significantly longer duration (Median=7 days, P<1e-16) compared to the majority younger sample (17-49 years) (Figure 2-d). Participants declared gender showed no significant difference in retention (P = 0.3). People with clinical conditions of interest to the study (e.g.: heart disease, depression, multiple sclerosis) remained in the studies for a significantly longer time (Median=13 days, P<1e-16) compared to participants that were recruited as non-disease controls(Median=6 days) (Figure 2-e). Median retention time also showed a marked and significant increase of 40 days (P<1e-16) for participants that were referred by a clinician to join one of the two studies (mPower and ElevateMS)(Median=44 days) compared to participants who self-selected to join the same study (Median=4 days) (Figure 2-f). See Supplementary tables 1-6 for a further breakdown of survival analysis results. Sensitivity analysis by including participants with missing age showed no impact on the association of age with participant retention. However, participants with missing demographics showed variation in retention compared to participants who shared their demographics(Supplementary Figure 1). This could be related to different time points at which demographic related questions were administered in individual studies.

**Participant Daily Engagement Patterns**

In the subgroup of participants who engaged with study apps for a minimum of 7 days, overall app usage clustered into four distinct groups with high (the dedicated cluster C1, and high utilizers in C2), moderate (cluster C3) and Sporadic (cluster C4) engagement (Figure 3.b). The participants who did not participate for at least 7 days were placed in a separate group of participants (the abandoners-C5*)(See Methods for cluster size determination and exclusion criteria details). The engagement and demographic characteristics across these five groups (C1-5*) varied significantly. Cluster 1 and 2 showed the highest daily app usage (Median app usage in the first 84 days = 96.4% and 63.1 % respectively) but also had the smallest proportion of participants (Median =9.5%) with the exception of Brighten where 23.7% of participants belonged to cluster C1. While daily app usage declined significantly for both moderate and sporadic clusters (C3- 21.4% and C4-22.6%), the median number of days between app usage was significantly higher for participants in the sporadic C4 cluster (Median=5 days) compared to cluster C3(Median=2 days). The majority of participants (median 54.6%) across the apps were linked to the abandoner group(C5*) with the median app usage of just 1 day(Figure 4.a-b). Furthermore, distinct demographic characteristics emerged across these five groups. Higher engagement clusters (C1-2) showed significant differences(P=1.38e-12) in proportion of adults 60 years and above (Median range

=15.1-17.2% across studies) compared to lower engagement clusters C3-5*(Median range =5.1-11.7% across studies)[Figure 4-c]. Minority groups such as Hispanic/Latinos, Asians, and African-American/Black, on the other hand, were represented in higher proportions in the clusters (C3-5*)(P=4.12e-10) with the least engagement(Figure 4-d] (See supplementary table 8 for further details).

**Discussion**

Our findings are based on one of the largest and diverse engagement dataset compiled to date. We identified two major challenges with remote data collection: (1) more than half of the participants discontinued participation within the first week of a study but that the rates at which people discontinued was drastically different based on age, disease status, clinical referral, and use of monetary incentives and (2) most studies were not able to recruit a representative sample, either demographically or regionally. Although these findings raise questions about the reliability and validity of data collected in this manner, they also shed light on potential solutions to overcome biases in populations using a combination of different recruitment and engagement strategies.

One solution could be the use of a flexible randomized withdrawal design[42]. Temporal retention analysis (Figure 2-c) shows that a run-in period could be introduced in the research design, wherein participants who are not active in the study app in the first week or two of the study can be excluded after enrollment but before the start of the actual study. The resulting smaller but more engaged cohort will help increase the statistical power of the study but does not fix the potential bias[43].

Another solution is to rely on monetary incentives to enhance engagement. Although only one study paid participants, the significant increase in retention and the largest proportion of frequent app users indicate the utility of the fair-share compensation model[1,44,45] in remote research. Such "pay-for-participation" model could be utilized by studies that require long-term and frequent remote participation. Researchers conducting case-control studies should also plan to further enrich and engage the population without the disease. Studies run the risk of not collecting sufficient data from controls to perform case-control analysis with participants without disease seen to be dropping out significantly early. Similarly, more efforts[46–48] are needed to retain the younger population that, although demonstrating large enrollment also features a majority dropping out on day one.

Distinct patterns in daily app usage behavior, also shown previously[49], further strengthen the evidence of unequal technology utilization in remote research. The majority of the participants found in the abandoners group (C5*) who dropped out of the study on day 1, may also reflect initial patterns in willingness to participate in research, in a way that cannot be captured by recruitment in traditional research. Put another way, although there is significant dropout in remote trials, these early drop-outs may be able to yield very useful information about differences in people who are willing to participate in research and those who are not willing to participate. For decades clinical research has been criticized for its potential bias because people who participate in research may be very different from people who do not participate in research[50–52]. Although researchers will not have longitudinal data from those who discontinue participation early, the information collected during onboarding can be used to assess potential biases in the final sample and may inform future targeted retention strategies.

Only 1 in 10 participants were in the high app use clusters (C1-2), and these clusters tended to be largely Non-Hispanic whites and older adults. Minority and younger populations, on the other hand, were represented more in the clusters with the lowest daily app usage (Figure 4-d). The largest impact on participant retention (>10 times) in the present sample was associated with clinician referral for participating in a remote study. This referral can be very light touch in nature, for example in the ElevateMS study, it consisted solely of clinicians handing patients a flyer with information about the study during a regular clinic visit. This finding is understandable, given recent research[53] showing that the majority of Americans trust medical doctors.

With the exception of Brighten study, the recruited sample was also inadequately diverse highlighting a persistent digital divide[54] and continued challenges in the recruitment of racial and ethnic underserved communities[55]. Additionally, the underrepresentation of States in the southern, rural and midwest regions indicates that areas of the US that often bear a disproportionate burden of disease[56] are under-represented in digital research[57,56,58]. This recruitment bias could impact future studies that aim to collect data for health conditions that are more prevalent among certain demographic[59] and associated with geographic groups[60]. Using different recruitment strategies[46–48] including targeted online ads in regions known to have a larger proportion of the minority groups, partnerships with local community organizations and clinics may help improve the penetration of remote research and improve diversity in the recruited sample. The ongoing "All of Us" research program that includes remote digital data collection has shown the feasibility of using a multifaceted approach to recruit a diverse sample with a majority of the cohort coming from communities underrepresented in biomedical research[61]. Additionally, simple techniques such as stratified recruitment that is customized based on the continual monitoring of the enrolling cohort demographics, can help enrich for a target population.

Finally, communication in digital health research may benefit from adopting the diffusion of innovations approach[62,63] that has been applied successfully in healthcare settings to change behavior including the adoption of new technologies[64–66]. Research study enrollments, advertisements including in-app communication and return of information to participants[67], could be tailored to fit three distinct personality types (trendsetters, majority, and laggards). While trendsetters will adopt innovations early, they are a minority (15%) compared to the majority (greater than two-thirds of the population) who will adopt a new behavior after hearing about its real-benefits, utility and believe it is the status quo. On the other hand, laggards (15%) are highly resistant to change and hard to reach online and as a result, will require more targeted and local outreach efforts.

These results should also be viewed within the context of limitations related to integrating diverse user-engagement data across digital health studies that targeted different disease areas with varying underlying disease characteristics and severity. We did not take into account differences in recruitment strategies used by the study apps. The present retention analysis is based on the "completed" tasks and did not account for incomplete tasks or time participants spent in the app. While sensitivity analysis showed the main findings from user retention analysis do not change by including participants with missing data, however, missing demographic characteristics remains to be a significant challenge for digital health(See supplementary table 7). Researchers should prioritize to collect minimal demographic data such as age, gender, race/ethnicity, participant state during onboarding which help characterize user attrition in future studies.

Despite these limitations, the present investigation to the best of our knowledge is the largest cross-study analysis of participant retention in remote digital health studies. While the technology has enabled researchers to reach and recruit participants for conducting large scale health research in short periods of time, more needs to be done to ensure equitable access and long-term utilization by participants across different populations. The low retention in "fully remote, app-based" health research may also need to be seen in the broad context of the mobile app industry where similar user attrition is reported[68]. Attrition in remote research may also be impacted by study burden[30] as frequent remote assessments can compete with users' everyday priorities and perceived value proposition for completing a study task that may not be linked to an immediate monetary incentive. Using co-design techniques[69] for developing study apps involving researchers and participants could help guide the development of most parsimonious research protocols that fit into the daily lives of people and are still sufficiently comprehensive for researchers.

In the present diverse sample of user-activity data, several cohort characteristics such as age, disease status, clinical referral, monetary benefits, etc have emerged as key drivers for higher retention. These characteristics may also guide the development of new data-driven engagement strategies[70,71] such as tailored just-in-time interventions[72] targeting sub-populations that are most likely to drop out early from remote research. Left unchecked the ongoing bias in participant recruitment combined with inequitable long-term participation in large scale "digital cohorts" can severely impact the generalizability[36,37] and undermine the promise of digital health in collecting representational real-world data.

**Online Methods**

**Data Acquisition**

The user engagement data was collected from eight digital health studies assessing different diseases ranging from parkinson's, asthma, heart condition, sleep health, multiple sclerosis to depression(Table 1). These studies recruited participants from throughout the US between 2014-2019 using a combination of different approaches including placing ads on social media, publicizing or launching the study at a large gathering, partnerships with patient advocacy groups, clinics, and through word of mouth.  The studies were launched at different time points during the 2014-2019 period, including three studies mPower, MyHeartCounts, and Asthma being launched with the public release of ResearchKit framework[73] released by Apple in March 2015. The studies were also active for different time periods including significant differences in the minimum time participants were expected to participate in the studies remotely. While Brighten and ElevateMS had a fixed 12 week participation period, other studies allowed participants to remain active for as long as they desired. Given this variation in the expected participation period across the studies, we selected the minimum common time period of the first 12 weeks(84 days) of each participant's activity in each study for retention analysis.  Finally, with the exception of Brighten study which was a randomized interventional clinical trial and enrolled depressed cohort offering them monetary incentives for participation, the rest of the seven studies were observational and did not offer any direct incentives for participation and were open to people with and without target disease. The studies also collected different real-world data ranging from frequent subjective assessments, objective sensor-based tasks to continual passive data[41] collection.

**Data Harmonization**

User activity data across all the apps were harmonized to allow for inter-app comparison of user engagement metrics. All in-app surveys and sensor-based tasks (eg. Finger tapping on the screen) were classified as "active tasks" data type. The data gathered without explicit user action such as daily step count (Apple's health kit API), daily local weather patterns were classified as "passive" data type and was not used for assessing active user engagement. The frequency at which the active tasks were administered in the study apps were aligned based on the information available in the corresponding study publication or obtained directly from the data contributing team in case the data was not publicly available. Furthermore, there were significant differences in the baseline demographics that were collected by each app. A minimal subset of four demographic characteristics (age, gender, race, state) was used for participant recruitment and retention analysis. A subset of six studies(mPower, ElevateMS, SleepHealth, Asthma, MyHeartCounts) had enrolled participants with and without disease status and were used to asses retention differences between people with(case) and without(control) disease. Two studies(mPower and ElevateMS) had a subset of participants that were referred to use the same study app by their care providers. For this smaller but unique sub-group, we compared the retention differences between clinically referred participants to self-referred participants.

**Statistical Analysis**

We used three key metrics to assess participant retention and long-term engagement. 1) Duration in the study: the total duration, a study participant remained active in the study i.e the number of days between the first and last active task completed by the participant, during the first 84 days of each participant's time in the study. 2) Days active in the study: the number of days a participant performed any active task in the app. 3) User activity streak: a binary-encoded vector representing the 84 days of app participation for each participant(Figure 3-a) where the position of the vector indicates the participant's day in the study and is set to 1(green box, Figure 3-a) if at least one active task is performed on that day or else is 0(white). User activity streak was used to assess sub-populations that show similar longitudinal engagement patterns over a 3 month period.

Participant retention analysis(survival analysis[74]) was done using the total duration in the study metric to compare the retention differences across studies, sex, age group, disease status, and clinical referral for study-app usage. Log-rank test[75] stratified by study type was used to compare significant differences in participant retention between different comparator groups. Kaplan-Meier[76] plots were used to summarize the effect of the main variable of interest by pooling the data across studies where applicable. Two approaches were used to evaluate participant retention using survival analysis. 1) No censoring(most conservative) - If the last active task completed by participant fell within the pre-specified study period of first 84 days, we considered it to be a true event i.e participant leaving the study (considered "dead" for survival analysis). b) Right-censoring[76] - To assess the sensitivity of our findings using approach 1, we relaxed the determination of true event (participant leaving the study) in the first 12 weeks to be based on the first 20 weeks of app activity (additional 8 weeks). For example, if a participant completes last task in an app on day 40(within the first 84 days) and then additionally completes more active task/s between week 13-20 he/she was still considered alive (no event) during the first 84 days(12 weeks) of the study and therefore "right-censored" for survival analysis.

Given that age and gender had a varying degree of missingness across studies; additional analysis comparing the retention differences between the two sub-groups that provided the demographics and that opted out was done to assess the sensitivity of missing data on main findings. Unsupervised k-means clustering was used to investigate the longitudinal participant engagement behavior within each study. The number of optimum clusters(between 1-10) in each study was determined using the elbow method[77] that aims to minimize the within-cluster variation. Enrichment of demographic characteristics in each cluster was assessed using a one-way analysis of variance(ANOVA). Since the goal of this unsupervised clustering of user activity streaks was to investigate the patterns in longitudinal participant engagement; we filtered out individuals who remained in the study for less than 7 days from clustering analysis. However, for post hoc comparisons of demographics across the clusters, the initially left-out users were put in a separate group (C5*). The state-wise proportions of recruited participants in each app were compared to the 2018 US state population estimates using the data obtained from the US census bureau[78]. To eliminate potential bias related to marketing and advertising of the launch of Apple's Research kit platform on March 09, 2015, participants who joined and left the mPower, MyHeartCounts, Asthma studies within the first week of Research Kit launch (N=15,413) were taken out from the user retention analysis. We initially considered using cox proportional hazards model[79] to test for the significance of variable of interest on user retention within each study accounting for other study-specific covariates. However, because the assumption of proportional hazards (tested using scaled Schoenfeld residuals) was not supported for some studies, these analyses were not further pursued. All statistical analyses were performed using R[80].

**Author Contributions**

AP conceptualized the study, wrote the initial analytical plan, carried out the analysis and wrote the first draft of the paper. AP and PS integrated and harmonized user engagement data across the studies. PAA and AP managed the data-sharing agreements with data contributors(NE, MM, CS, and DG). ECN, PH, LO made significant contributions to the data analysis. PAA, LM, and LO helped interpret the analytical findings and provided feedback on the initial manuscript draft. AP and PAA made major revisions to the first draft. CS and JW implemented the data governance and sharing model to enable the sharing of raw user engagement data with the research community. LO and PAA jointly oversaw the study. All authors assisted with the revisions of the paper.

**Acknowledgments**

Funding Support: This work was supported in part by various funding agencies that included the National Institute of Mental Health (MH100466), Robert Wood Johnson Foundation (RWJF - 73205), National Library of Medicine (R01LM013043), National Center for Advancing Translational Sciences (1UL1TR002319-01) and American Sleep Apnea Foundation, Washington, DC. The funding agencies did not play a role in study design; in the collection, analysis and interpretation of data; in the writing of the report; or in the decision to submit this article for publication. We also would like to thank Phendo[81], SleepHealth[82], and GoodRx[83] groups for sharing user engagement data from their respective digital health studies for the retention analysis. We also acknowledge and thank researchers from previously completed digital health studies[16,23,84–86] for making the user-engagement data available to the research community under qualified researcher program[87]. Specifically, mPower study data were contributed by users of the Parkinson mPower mobile application as part of the mPower study developed by Sage Bionetworks and described in Synapse [doi:10.7303/syn4993293]. MyHeartCounts study data were contributed by users of the My Heart Counts Mobile Health Application study developed by Stanford University and described in Synapse [doi:10.7303/syn11269541]. Asthma study data were contributed by users of the asthma app as part of the Asthma Mobile Health Application study developed by The Icahn School of Medicine at Mount Sinai and described in Synapse [doi:10.7303/syn8361748]. We would like to thank and acknowledge all study participants for contributing their time and effort to participate in these studies.

# Figure & Tables

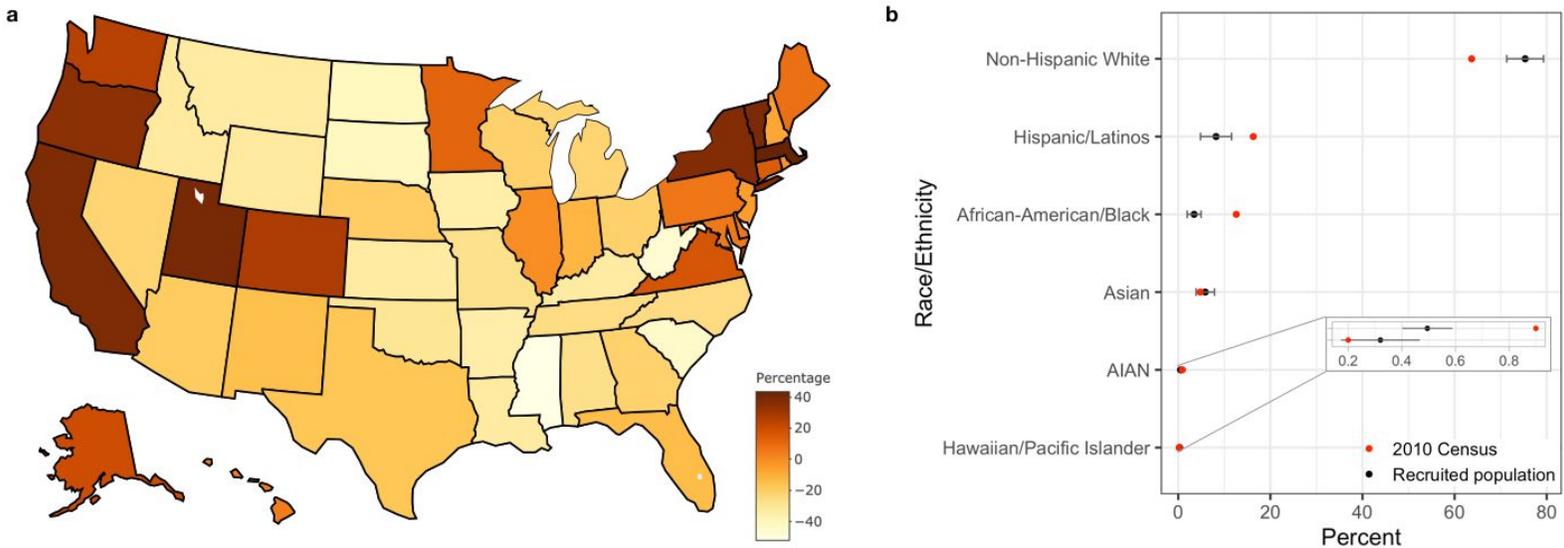

**Figure 1. a)** Map of US showing the ratio of the percentage of recruited participants to state's population proportion of the US(median across the studies) and **b)** Race/Ethnicity proportion(median +/- IQR) of recruited participants compared to 2010 census data.

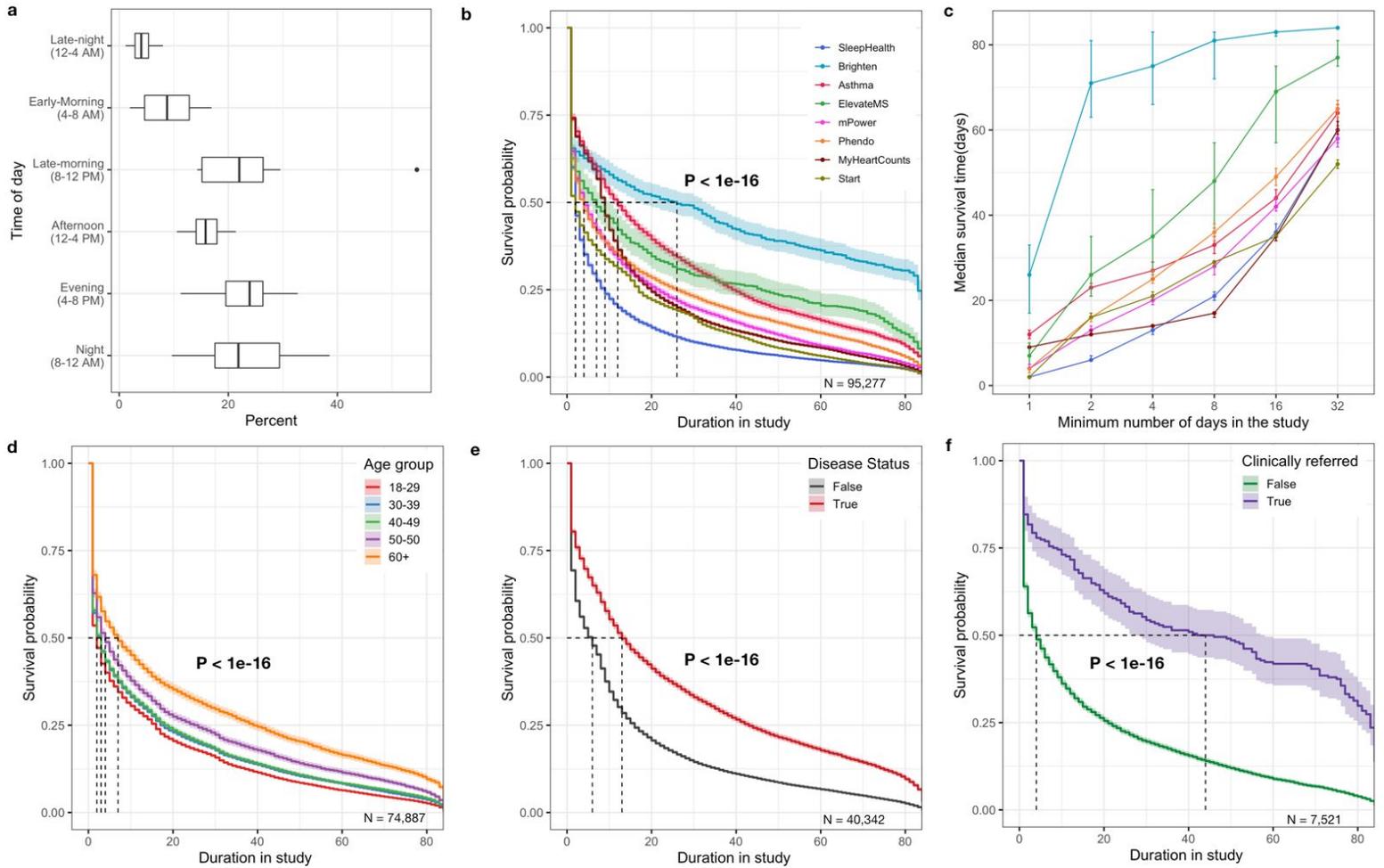

**Figure 2: a)** Proportion of active tasks (N = 3.3 million) completed by participants based on their local time of day, **b)** Kaplan Meir survival curve showing significant differences(P < 1e-16) in user retention across the apps. Brighten App where monetary incentives were given to participants showed the longest retention time(Median = 26 days, 95% CI= 17-33) followed by Asthma(Median = 12 days, 95% CI= 11-13), MyHeartCounts(Median = 9 days, 95% CI= 9-9), ElevateMS(Median = 7 days, 95% CI= 5-10), mPower(Median = 5 days, 95% CI= 4-5), Phendo(Median = 4 days, 95% CI= 3-4), Start(Median = 2 days, 95% CI= 2-2) and SleepHealth(Median = 2 days, 95% CI= 2-2), **c)** Lift curve showing the change in median survival time (with 95% CI) based on the minimum number of days(1-32) a subset of participants continued to use the study apps, Kaplan-Meier survival curve showing significant differences in user retention across **d)** Age group, with 60 years and older using the apps for longest duration(Median=7days, 95%CI=6-8, P < 1e-16) followed by 50-59 years (Median =4 days, 95%CI= 4-5) and 17-49 years (Median = 2-3 days, 95% CI= 2-3) **e)** Disease status; participants reporting having a disease stayed active longer($N_{50}$= 13days, 95% CI=13-14) compared to people without disease($N_{50}$= 6 days, 95% CI=5-6) and finally **f)** Clinical referral; Two studies (mPower and ElevateMS), had a subpopulation, that were referred to the study by clinicians and showed significantly(P<1e-16) longer

app usage period(Median= 44 days, 95% CI=27-58) compared to self-referred participants with disease ($N_{50}$= 4 days, 95% CI=4-4).

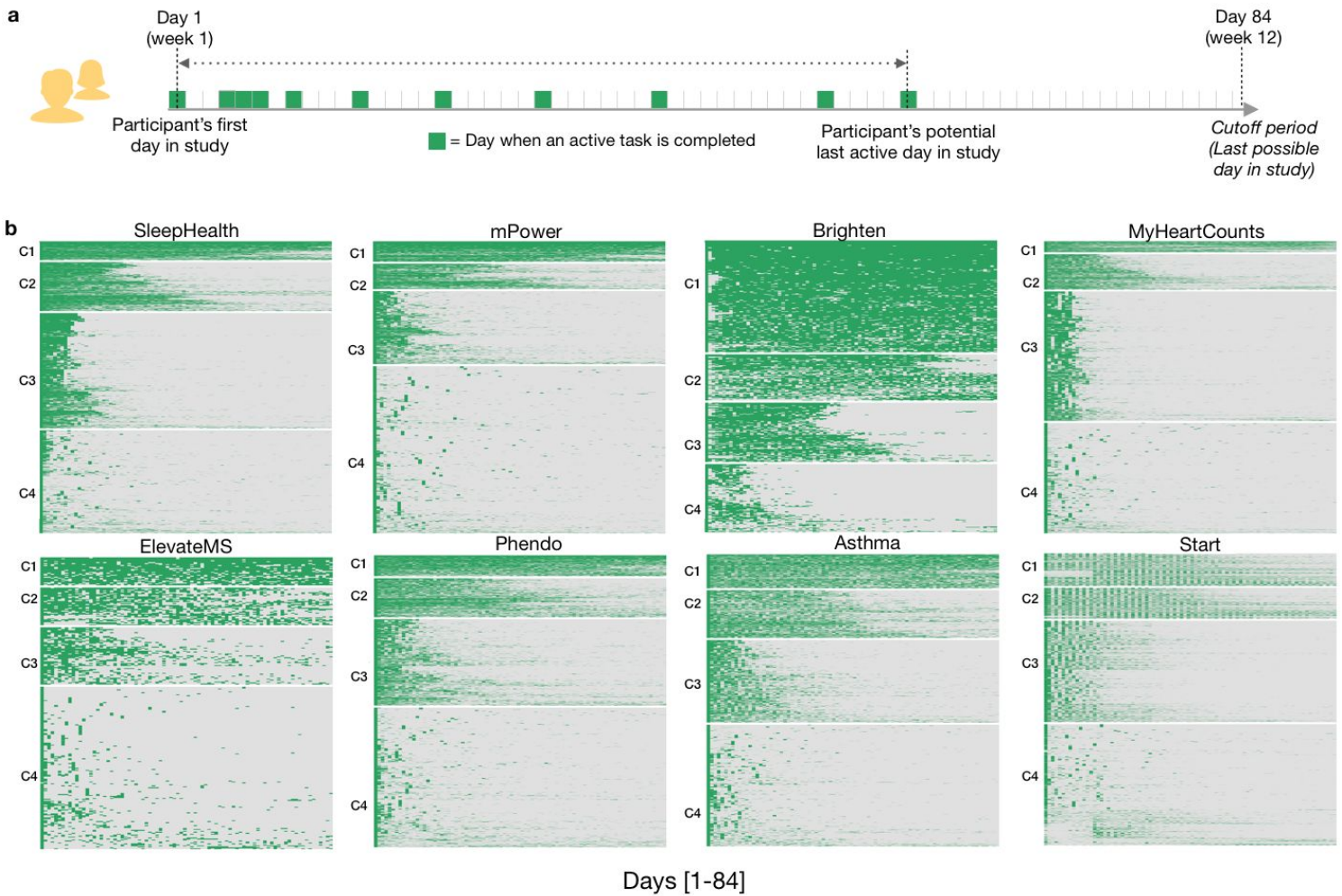

**Figure 3: a)** Schematic representation of an individual's in-app activity for the first 84 days. The participant app usage time is determined based on the number of days between the first and last day they perform an active task(indicated by the green box) in the app. Days active in the study is the total number of days a participant performs at least one active task (indicated by the number of green boxes). **b)** Heatmaps showing participants in-app activity across the apps for the first 12 weeks(84 days), grouped into four broad clusters using unsupervised k-means clustering. The optimum number of clusters was determined by minimizing the within-cluster variation across different cluster sizes between 1-10. Seven out of eight studies indicated four clusters to be an optimum number using the elbow method. The heatmaps are arranged by the highest (C1) to the lowest user engagement cluster (C4).

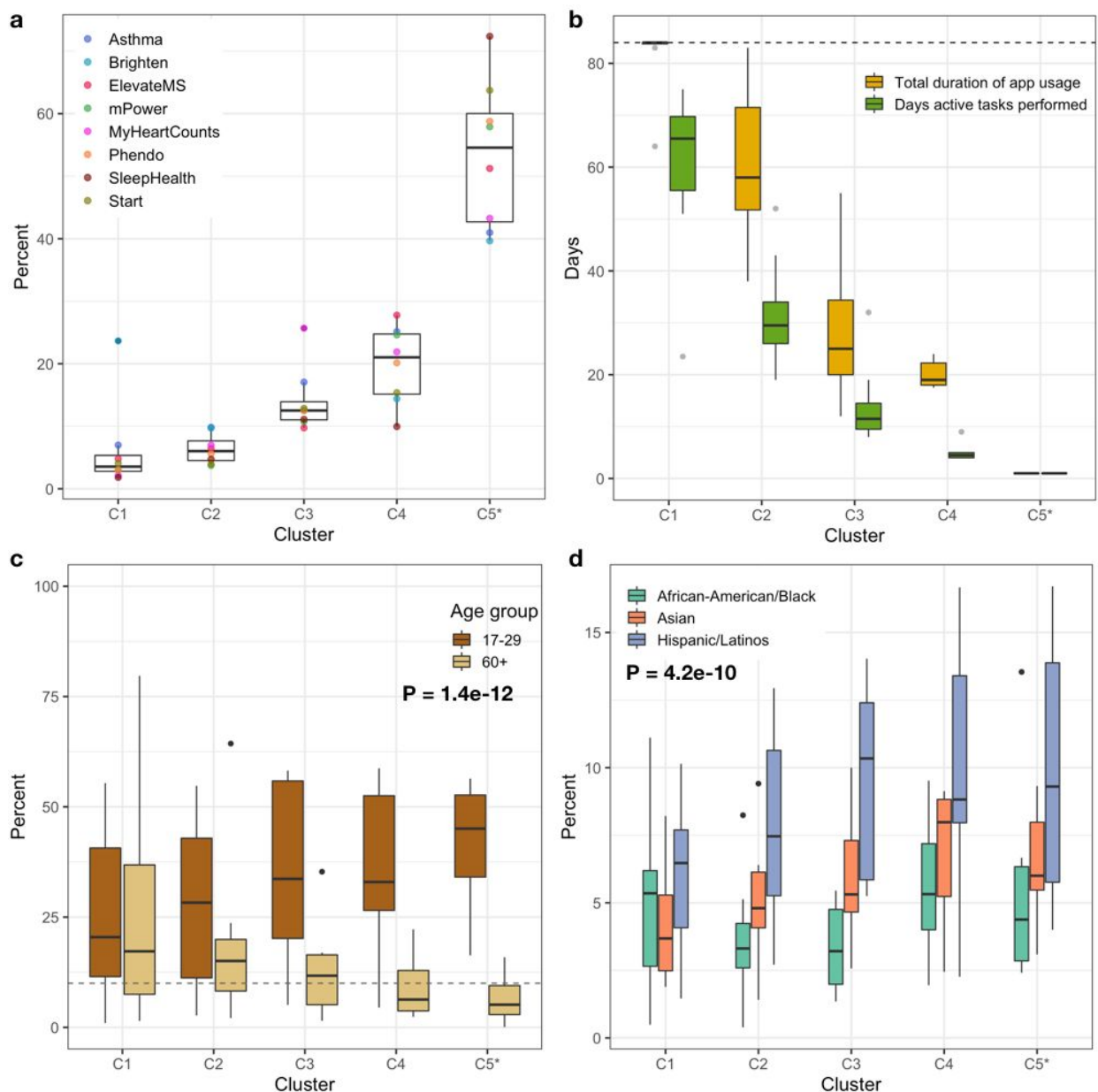

**Figure 4: a)** Proportion of participants in each cluster across the study apps, **b)** Participants total app usage duration(between 1-84 days) and the number of days participants completed tasks in the study apps, **c)** Significant differences[F(4,163)=18.5, P=1.38e-12] in proportion of participants aged 17-29 years and 60 years and older across the 5 clusters and **d)** Significant differences[F(2,81)=28.5, P=4.12e-10] in proportion of minority population present in the five clusters. C5* cluster contains the participants that used the apps for less than a week and were removed from the clustering, however, they were added back for accurate proportional comparison of participants in each cluster.

| Study | Disease Focus / Study type | Study period | Number of participants | Total participant days | Active tasks completed |
|---|---|---|---|---|---|
| Start | Antidepressant Efficacy - Observational | Aug,2015 - Feb,2018 | 42,704 | 280,489 | 1,219,656 |
| MyHeartCounts | Cardiovascular Health - Observational | Mar,2015 - Oct,2015 | 26,902 | 165,455 | 305,821 |
| SleepHealth | Sleep Apnea - Observational | Jul,2015 - Jun,2019 | 12,914 | 99,696 | 401,628 |
| mPower | Parkinson's - Observational | Mar,2015 - Jun,2019 | 12,236 | 104,797 | 568,685 |
| Phendo | Endometriosis - Observational | Dec,2016 - Jul,2019 | 7,802 | 81,938 | 735,778 |
| Asthma | Asthma - Observational | Mar,2015 - Dec,2016 | 5,875 | 77,815 | 175,699 |
| Brighten | Depression - Randomized Control Trial | Jul,2014 - Aug,2015 | 876 | 34,987 | 45,951 |
| ElevateMS | Multiple Sclerosis - Observational | Aug,2017 - Jul,2019 | 605 | 11,211 | 31,568 |
| | | | **109,914** | **856,388** | **3,484,786** |

**Table 1: Summary of user engagement data compiled from eight digital health studies**

|  | Asthma | Brighten | ElevateMS | mPower | MyHeart Counts | Phendo | SleepHealth | Start | Overall (median) |
|---|---|---|---|---|---|---|---|---|---|
| **Age group** | | | | | | | | | |
| *N* | *2512* | *875* | *569* | *6810* | *1555* | *7484* | *12392* | *42690* | |
| 18-29(%) | 43.31 | 50.06 | 10.9 | 31.5 | 25.08 | 55.38 | 32.79 | 55.72 | 38 |
| 30-39(%) | 27.83 | 25.14 | 26.54 | 18.37 | 32.67 | 36.09 | 28.72 | 24.14 | 27.2 |
| 40-49(%) | 14.41 | 14.74 | 28.47 | 13.19 | 16.27 | 8.23 | 20.77 | 12.38 | 14.6 |
| 50-59(%) | 9.08 | 6.97 | 22.14 | 13.61 | 12.09 | 0.25 | 11 | 5.26 | 10 |
| 60+(%) | 5.37 | 3.09 | 11.95 | 23.33 | 13.89 | 0.04 | 6.72 | 2.51 | 6 |
| **Sex** | | | | | | | | | |
| *N* | *2509* | *875* | *329* | *6916* | *6976* | *7532* | *12558* | *42704* | |
| Female(%) | 39.58 | 77.83 | 74.16 | 28.93 | 18.94 | 100 | 29.14 | 75.86 | 56.9 |
| **Race** | | | | | | | | | |
| *N* | *3274* | *875* | *334* | *6884* | *4703* | *7530* | *5311* | *-* | |
| Non-Hispanic White(%) | 68.69 | 60.11 | 80.84 | 75.32 | 77.95 | 81.29 | 74.13 | - | 75.3 |
| Hispanic/Latinos(%) | 13.29 | 14.29 | 4.79 | 8.21 | 6.97 | 5.67 | 12.82 | - | 8.21 |
| African-American/Black(%) | 4.95 | 10.86 | 6.89 | 2.05 | 3.1 | 2.71 | 3.45 | - | 3.45 |
| Asian(%) | 4.98 | 8.23 | 2.99 | 8.4 | 7.72 | 2.79 | 5.87 | - | 5.9 |
| Hawaiian or other Pacific Islander(%) | 0.89 | 0.57 | 0 | 0.28 | 0.32 | 0 | 0.23 | - | 0.3 |
| AIAN(%) | 0.43 | 0.46 | 0 | 0.65 | 0.53 | 0.74 | 0.28 | - | 0.5 |
| Other(%) | 6.78 | 5.49 | 4.49 | 5.1 | 3.4 | 6.8 | 3.22 | - | 5.1 |
| **Duration in Study** (Median +/- IQR) | 12 ± 38 | 26 ± 82 | 7 ± 45 | 4 ± 21 | 9 ± 19 | 4 ± 25 | 2 ± 8 | 2 ± 16 | 5.5 |
| **Days active tasks performed** (Median +/- IQR) | 4 ± 12 | 14 ± 58 | 2 ± 8 | 2 ± 4 | 4 ± 7 | 2 ± 6 | 2 ± 4 | 2 ± 4 | 2 |

**Table 2: Summary of select participant demographics and study app usage across the eight digital health studies**